\documentclass[12pt]{iopart}
\usepackage{epsf,graphics,graphicx}
\usepackage{iopams}
\def\n{\nonumber}

\def\ket{\rangle}
\def\bra{\langle}
\def\Tr{{\textrm{Tr}}}
\begin{document}
\title{Berry phase and fidelity susceptibility of the three-qubit
Lipkin-Meshkov-Glick ground state}
\author{Erik Sj\"oqvist}
\address{Department of Quantum
Chemistry, Uppsala University, Box 518, Se-751 20 Sweden}
\ead{eriks@kvac.uu.se}
\author{Ramij Rahaman}
\address{Selmer Center,
Department of Informatics, University of Bergen, Bergen, N5020,
Norway}
\author{Urna Basu}
\address{Theoretical Condensed Matter Physics Division,
Saha Institute of Nuclear Physics, Kolkata 700064, India}
\author{B. Basu}
\address{Physics and Applied Mathematics Unit,
Indian Statistical Institute, Kolkata 700108, India}
\ead{banasri@isical.ac.in}
\date{\today}
\begin{abstract}
Berry phases and quantum fidelities for interacting spins have attracted
considerable attention, in particular in relation to entanglement properties
of spin systems and quantum phase transitions. These efforts mainly
focus either on spin pairs or the thermodynamic infinite spin limit, while
studies of the multipartite case of a finite number of spins are rare.
Here, we analyze Berry phases and quantum fidelities of the energetic ground
state of a Lipkin-Meshkov-Glick (LMG) model consisting of three
spin-$\frac{1}{2}$ particles (qubits). We find explicit expressions
for the Berry phase and fidelity susceptibility of the full system
as well as the mixed state Berry phase and partial-state fidelity
susceptibility of its one- and two-qubit subsystems. We demonstrate
a realization of a nontrivial magnetic monopole structure associated
with local, coordinated rotations of the three-qubit system around
the external magnetic field.
\end{abstract}
\submitto{\JPA}
\maketitle

\section{Introduction}
Recently, the study of various ground state properties of many-body
systems has attracted a lot of attention in the context of quantum phase
transition (QPT). QPTs, driven solely by quantum fluctuations, exhibit a
dramatic change in the ground state at zero temperature under change of external
parameters and is associated with a level crossing or avoided crossing between
the ground state and first excited state \cite{sachdev93}. A deeper
understanding of QPTs has emerged from the fundamentals of quantum mechanics.
In particular, when the external parameters of the Hamiltonian are varied,
the response of the energetic ground state of the
system has been analyzed in terms of Berry phase \cite{berry84} and
Bures-Uhlmann fidelities \cite{bures69,uhlmann76}. These quantities show nontrivial
behavior related to points in parameter space where two or more energy levels
become degenerate.

For spin-like systems, energy crossings provide realizations of
magnetic monopole structures; a fact that has opened up for simulations of
magnetic monopoles in the laboratory \cite{fan03}. These monopoles give rise
to a magnetic flux that can be measured as a Berry phase being proportional to
the area (solid angle) enclosed in the parameter space of the system. The monopole structure
of interacting spins has been examined in several studies in the past
\cite{yi04,sjoqvist05,kwan08,oh09a,oh09b,niu10}.

The Berry phase for systems with several spin$-\frac{1}{2}$ (qubits) has
been addressed recently. Lei Xing \cite{xing06} examined a three-qubit
model with uniaxial qubit-qubit interaction and demonstrated that the
corresponding Berry phase admits a solid angle interpretation, provided
the couplings are added to the underlying parameter space. Williamson and
Vedral \cite{williamson07,williamson09} found a nontrivial relation between the Berry
phase of translationally symmetric multi-qubit states and their multi-partite
entanglement properties. The behavior of the Berry phase in the thermodynamic
limit has been studied for $XY$ spin-chains \cite{carollo05,zhu06,basu07,basu10},
the Dicke model \cite{chen06}, and the Lipkin-Meskov-Glick
(LMG) system \cite{cui06}.  In this paper, we analyze the Berry phase
for an analytically solvable, finite-size LMG type model \cite{lipkin65}
consisting of three spin$-\frac{1}{2}$ particles. We examine the Berry phase
structure of the energetic ground state of the LMG system and its subsystems
using pure state \cite{berry84} and mixed state \cite{sjoqvist00b} Berry phases.

The fidelity \cite{zanardi06} is an information
theoretic measure that can be used to analyze the quantal properties of
the ground state of spin systems. The utility of this measure and the
related fidelity susceptibility, has been explored in a number of studies
\cite{you07,chen08,tribedi08}, in particular in relation to QPTs.
The fidelity susceptibility and the Berry phase are two complementary
manifestations of the underlying geometry of the state space, as described
by the quantum geometric tensor \cite{zanardi07,venuti07}. Recently, the
concept of partial-state fidelity has been developed, which measures
the fidelity of a subsystem along with the associated notion of
partial-state fidelity susceptibility \cite{paunkovic08a,kwok08}. Here,
we wish to examine the response of the three-qubit LMG ground state to
parameter variation in terms of the fidelity susceptibility and
partial-state fidelity susceptibility.

The outline of the paper is as follows. In the next section, the
three-qubit LMG model is described, the corresponding energy eigenvalues
and eigenvectors are found, and the ground states are identified.
Secs. \ref{sec:berry} and \ref{sec:fidelity} examine in detail the Berry
phases and fidelity susceptibilities of the present LMG system and its
subsystems. The paper ends with the conclusions.

\section{Three-qubit LMG model}
The LMG model of spin systems has found applications in Bose-Einstein
condensates \cite{cirac98}, statistical mechanics of mutually interacting
spins \cite{ribeiro07}, and entanglement theory \cite{latorre05,orus08}.
The LMG model describes a set of $N$ qubits (spin-$\frac{1}{2}$) mutually
interacting through a $XY$-like term in the Hamiltonian and coupled to an
external transverse magnetic field. The ferromagnetic version of the LMG
Hamiltonian reads
\begin{eqnarray}
H(\gamma,h;N) = -\frac{1}{N} (S_x^2+\gamma S_y^2) -h S_z ,
\end{eqnarray}
where $\gamma$ is an anisotropy parameter ($\gamma = 1$ corresponds to
the isotropic LMG model), $h$ is the strength of an
external magnetic field in the $z$ direction, and $S_\alpha=\sum_{k=1}^N
\frac{1}{2} \sigma_\alpha^k$ is the $\alpha$th component of
the total spin operator ($\hbar = 1$ from now on) with $\sigma_x^k =
|0 \ket\bra 1| + |1 \ket\bra 0|, \sigma_y^k = -i|0 \ket\bra 1| +
i|1 \ket\bra 0|$, and $\sigma_z^k = |0 \ket\bra 0| - |1 \ket\bra 1|$
the Pauli operators of the $k$th qubit. In the present work, we examine
the exactly solvable three-qubit ($N=3$) case.

The Hamiltonian for the three-qubit LMG system is given by
\begin{eqnarray}
H(\gamma,h) & = & H(\gamma,h;3) =
-\frac{1}{6} \left[ \sigma_x^1\sigma_x^2+ \sigma_x^2\sigma_x^3 +
\sigma_x^1\sigma_x^3 + \gamma (\sigma_y^1\sigma_y^2+\sigma_y^2\sigma_y^3 +
\sigma_y^1\sigma_y^3) \right]
\n \\
 & & -\frac h2 \left( \sigma_z^1+\sigma_z^2+\sigma_z^3 \right) ,
\end{eqnarray}
where we have ignored the unimportant constant term $-\frac{1}{4} (1+\gamma)$.
In the computational basis $\{ |000\ket,|011\ket,|101\ket ,|110\ket ,
|111\ket, |100\ket , |010\ket,|001\ket \}$, the Hamiltonian takes the
block-diagonal form
\begin{eqnarray}
\boldsymbol{H} (\gamma,h) = \left( \begin{array}{cc}
\boldsymbol{P} (\gamma,h) & \boldsymbol{0} \\
\boldsymbol{0} & \boldsymbol{P} (\gamma,-h)
\end{array} \right) ,
\end{eqnarray}
where
\begin{eqnarray}
\boldsymbol{P} (\gamma,h) & = &
\left( \begin{array}{cccc}
\frac{3}{2}h & -\frac{1}{6}(1-\gamma) &
-\frac{1}{6}(1-\gamma) & -\frac{1}{6}(1-\gamma) \\
-\frac{1}{6}(1-\gamma) & -\frac{h}{2} &
-\frac{1}{6}(1+\gamma) & -\frac{1}{6}(1+\gamma) \\
-\frac{1}{6}(1-\gamma) & -\frac{1}{6}(1+\gamma) &
-\frac{h}{2} & -\frac{1}{6}(1+\gamma) \\
-\frac{1}{6}(1-\gamma) & -\frac{1}{6}(1+\gamma) &
-\frac{1}{6}(1+\gamma) & -\frac{h}{2}
\end{array} \right) ,
\end{eqnarray}
and $\boldsymbol{0}$ is the $4\times 4$ null matrix. The block structure
originate from the existence of the conserved quantity $\sigma_1^1
\sigma_1^2 \sigma_1^3$, which displays the fact that only states with
same spin parity interact \cite{dusuel04}. Due to the
$h \leftrightarrow -h$ symmetry between the two $\boldsymbol{P}$
blocks of $\boldsymbol{H}$, we may assume $h\geq 0$ without loss of
generality. To facilitate
the diagonalization of $\boldsymbol{H}$, we define energy functions
$\mathcal{E}_{\pm}$ and $\mathcal{E}$, and mixing angle $\Theta$
according to
\begin{eqnarray}
\mathcal{E}_{\pm} (\gamma,h) & = &
\frac{1}{6}\left( 3h-1-\gamma \pm 2 \sqrt{9h^2+3h(1+\gamma) +
1-\gamma +\gamma^2} \right)
\nonumber \\
 & = & \mathcal{E}_0 (\gamma,h) \pm \Delta \mathcal{E} (\gamma,h),
\nonumber \\
\tan \left[ \frac{1}{2} \Theta (\gamma,h) \right] & = &
\frac{\sqrt{3}(\gamma -1)}{6h + 1+\gamma -
2 \sqrt{9h^2+3h(1+\gamma) + 1-\gamma +\gamma^2}} .
\end{eqnarray}
Note, in particular, that $\tan \left[ \frac{1}{2} \Theta (\gamma,h) \right]$
($\tan \left[ \frac{1}{2} \Theta (\gamma,-h) \right]$) diverges (tends to
zero) in the isotropic limit $\gamma \rightarrow 1$. Thus, $\Theta (1,h) = \pi$
and $\Theta (1,-h) = 0$. We may now write the eigenvalues $E$ and orthonormalized
eigenvectors $|V\ket$ of $H$ in terms of $\mathcal{E}_{\pm},\mathcal{E}$, and
$\Theta$ as
\begin{eqnarray}
E_{+}^{(+)} & = & \mathcal{E}_{+} (\gamma,h) : \ |V_{+}^{(+)} \ket =
-\sin \left[ \frac{1}{2} \Theta (\gamma,h)
\right] |000\ket +
\cos \left[ \frac{1}{2} \Theta (\gamma,h) \right]
|\overline{W}\ket ,
\nonumber \\
E_{-}^{(+)} & = & \mathcal{E}_{-} (\gamma,h) : \ |V_{-}^{(+)} \ket =
\cos \left[ \frac{1}{2} \Theta (\gamma,h)
\right] |000\ket +
\sin \left[ \frac{1}{2} \Theta (\gamma,h) \right]
|\overline{W}\ket ,
\nonumber \\
E^{(+)} & = & -\mathcal{E}_0 (\gamma,h) : \
\left\{ \begin{array}{l}
|V_{1}^{(+)} \ket = \frac{1}{\sqrt{2}} (|011 \ket - |110 \ket ) , \\
|V_{2}^{(+)} \ket = \frac{1}{\sqrt{6}} (|011 \ket -2 |101 \ket +
|110 \ket ) ,
\end{array} \right.
\nonumber \\
E_{+}^{(-)} & = & \mathcal{E}_{+} (\gamma,-h) : \ |V_{+}^{(-)} \ket =
-\sin \left[ \frac{1}{2} \Theta (\gamma,-h)
\right] |111\ket +
\cos \left[ \frac{1}{2} \Theta (\gamma,-h) \right]
|W\ket ,
\nonumber \\
E_{-}^{(-)} & = & \mathcal{E}_{-} (\gamma,-h) : \ |V_{-}^{(-)} \ket =
\cos \left[ \frac{1}{2} \Theta (\gamma,-h)
\right] |111\ket +
\sin \left[ \frac{1}{2} \Theta (\gamma,-h) \right]
|W\ket ,
\nonumber \\
E^{(-)} & = & - \mathcal{E}_0 (\gamma,-h) : \
\left\{ \begin{array}{l}
|V_{1}^{(-)} \ket = \frac{1}{\sqrt{2}} (|100 \ket - |001 \ket ) , \\
|V_{2}^{(-)} \ket = \frac{1}{\sqrt{6}} (|100 \ket -2 |010 \ket +
|001 \ket ) ,
\end{array} \right.
\label{eq:eigensolutions}
\end{eqnarray}
where $|\overline{W} \ket = \sigma_x \otimes \sigma_x \otimes \sigma_x
|W \ket = \frac{1}{\sqrt{3}} (|011\ket + |101\ket +|110\ket)$.

Alternatively, we may use that the total spin $S^2$ commutes with
the LMG Hamiltonian \cite{dusuel04}, which implies that the eigensolutions
may be labeled by the total spin. For instance, we may write the two
types of ground states as $|V_{-}^{(\pm)} \ket = \cos \left[ \frac{1}{2}
\Theta (\gamma,\pm h) \right] |\frac{3}{2},\pm\frac{3}{2} \ket +
\sin \left[ \frac{1}{2} \Theta (\gamma,\pm h) \right]
|\frac{3}{2}, \mp \frac{1}{2} \ket$ with $|S,M\ket$ being the
common eigenvectors of $S^2$ and $S_z$.

\begin{figure}
\includegraphics[width=8cm]{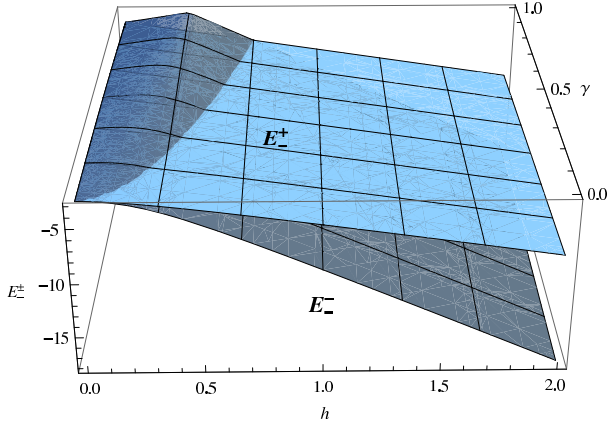}
\caption{\label{fig:E_cross} Energy levels $E_-^{(+)}$ and
$E_+^{(-)}$ of the two
potential ground states $V_{-}^{(+)}$ and $V_{-}^{(-)}$ as a function
of the isotropy parameter $\gamma$ and the magnetic field strength $h$.
(This figure is in colour only in the electronic version)}
\end{figure}

The lowest energy state is $V_{-}^{(+)}$ or $V_{-}^{(-)}$
associated with energies $E_{-}^{(+)}$ and $E_{-}^{(-)}$,
respectively, shown in
Fig.\ref{fig:E_cross}. The crossing points lie along
the lines $\gamma \mapsto h_c = h_c (\gamma)$, which are determined by
\begin{eqnarray}
\mathcal{E}_{-} (\gamma,h_c) = \mathcal{E}_{-} (\gamma,-h_c) .
\end{eqnarray}
This yields the following two classes of solutions
\begin{eqnarray}
h_c^{(1)}  & = & 0,  \ \gamma \ {\textrm{arbitrary}},
\end{eqnarray}
and
\begin{eqnarray}
(h_c^{(2)})^2 & = & \frac{4}{9} \gamma .
\end{eqnarray}
The energetic ground state in the low field strength regime, corresponding
to $h_c^{(1)} < h<h_c^{(2)}$, is $V_{-}^{(+)}$; in the high field strength
regime, corresponding to $h>h_c^{(2)}$, it is $V_{-}^{(-)}$.

The energetic ground state may take any of the three main three-qubit forms:
$W$, GHZ, and product states. Here, we identify the corresponding mixing angles
and delineate the exact form of these ground states. The $W$ and product forms
are obtained for mixing angle being an integer multiple of $\pi$, which may happen
only in the isotropic case $\gamma = 1$. Indeed, we found above that
$\Theta (1,h)=\pi$ and $\Theta (1,-h)=0$, which implies the energetic
ground state
\begin{eqnarray}
|V_{-}^{(+)} \ket & = & |\overline{W} \ket, \ \ h_c^{(1)} < h < h_c^{(2)} ,
\nonumber \\
|V_{-}^{(-)} \ket & = & |111 \ket, \ \ h > h_c^{(2)}
\label{eq:gsisotropic}
\end{eqnarray}
in the isotropic LMG model. The GHZ form $U_1 \otimes U_2
\otimes U_3 \frac{1}{\sqrt{2}} (|000\ket \pm |111\ket)$, $U_1,U_2,U_3$ any
one-qubit unitary operators, requires mixing angles that satisfy
$\tan \left( \frac{1}{2}\Theta \right) = \pm \sqrt{3}$. This happens at
$\Theta = \Theta (0,0) = \frac{2\pi}{3}$ and $\Theta =
\Theta (2,-\frac{1}{3}) = \frac{4\pi}{3}$. Here, $\Theta = \frac{2\pi}{3}$
corresponds to the two-fold degenerate ground state
\begin{eqnarray}
|V_{-}^{(\pm)} \ket & = &
\sigma_x \otimes \sigma_x \otimes \sigma_x
\frac{1}{\sqrt{2}} (|000\ket \pm |111\ket) ,
\label{eq:gsorigin}
\end{eqnarray}
i.e., GHZ form with $U_1=U_2=U_3=\sigma_x$. Thus, the energetic ground state
tends to a GHZ when approaching the origin in the $(\gamma,h)$ plane. The
angle $\Theta = \frac{4\pi}{3}$ yields a GHZ with $U_1=U_2=U_3=\sigma_y$.
However, the corresponding states at $(\gamma,h)=(2,\pm \frac{1}{3})$ are
$V_{-}^{(\mp)}$, neither of which being the energetic ground state. In
other words, the ground state may be of GHZ form only at $(\gamma,h)=(0,0)$.

\section{Berry phase}
\label{sec:berry}

\subsection{Full system}

\begin{figure}[b]
\includegraphics[width=7cm]{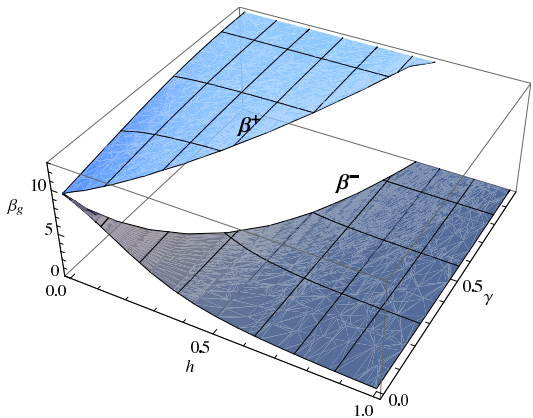}
\caption{\label{fig:bp}
The absolute value $|\beta_g|$ of the ground state Berry phase $\beta_g$ as a
function of the isotropy parameter $\gamma$ and the magnetic field strength $h$.
(This figure is in colour only in the electronic version)}
\end{figure}

Here, we examine Berry phases \cite{berry84} arising in adiabatic variation
of the LMG Hamiltonian. For given $\gamma$ and $h$, let us consider the
isospectral one-parameter Hamiltonian family
\begin{eqnarray}
H(\gamma,h;\phi) =
e^{-i\phi S_z} H(\gamma,h) e^{i\phi S_z} ,
\end{eqnarray}
where $\phi$ is slowly varying. Note that the unitary operator
$e^{-i\phi S_z}$, corresponding to coordinated rotation of the system
around the $z$ axis by an angle $\phi$, preserves the $4 \times 4$
block-structure of $H(\gamma,h)$. Thus, by preparing the system in the
energetic ground state and by varying $\phi$ slowly, the system remains
in the corresponding two-dimensional subspace. The state of the system may
be represented by one of the double-valued eigenvectors
\begin{eqnarray}
|V_{-}^{(\pm)} (\phi) \ket = e^{-i\phi S_z} |V_{-}^{(\pm)} \ket .
\end{eqnarray}
After completion of a $2\pi$ rotation around the $z$ axis, corresponding
to increasing $\phi$ from $0$ to $2\pi$, we obtain the Berry phase in
cyclic adiabatic evolution as \cite{mukunda93,polavieja98}
\begin{eqnarray}
\beta_g = \left\{ \begin{array}{lll}
\beta^{(+)} & = & \arg \bra V_{-}^{(+)} (0) | V_{-}^{(+)} (2\pi) \ket +
i\int_0^{2\pi} \langle V_{-}^{(+)} (\phi)|\frac{\partial}{\partial \phi}
V_{-}^{(+)} (\phi) \ket d \phi \\
 & & \\
 & = & -2\pi \left( 1-\cos \left[ \Theta (\gamma,h) \right] \right) , \ \
h_c^{(1)} <  h<h_c^{(2)} \\
 & & \\
\beta^{(-)} & = & \arg \bra V_{-}^{(-)} (0) | V_{-}^{(-)} (2\pi) \ket +
i\int_0^{2\pi} \langle V_{-}^{(-)} (\phi)|\frac{\partial}{\partial \phi}
V_{-}^{(-)} (\phi) \ket d \phi \\
 & & \\
 & = & 2\pi \left( 1-\cos \left[ \Theta (\gamma,-h) \right] \right) , \ \
h>h_c^{(2)}
\end{array} \right.
\label{eq:purebp}
\end{eqnarray}
The absolute value $|\beta_g|$ of the ground state Berry phase $\beta_g$ is shown in
Fig. \ref{fig:bp}. It should be noted that the Berry phase is defined modulus
$2\pi$, which implies that the $4\pi$ jump at the crossing point $h_c^{(2)} =
\frac{2}{3}$ in the isotropic ($\gamma = 1$) LMG model that is visible in Fig.
\ref{fig:bp} cannot be detected experimentally.

In order to understand the origin of the nontrivial two-level type $\beta_g$
in Eq. (\ref{eq:purebp}), we project the Hamiltonian $H(\gamma,h;\phi)$ onto
two-dimensional subspaces spanned by $\{ |000\ket,
|\overline{W}\ket \}$ for $h_c^{(1)} < h < h_c^{(2)}$ and $\{ |111\ket,|W\ket \}$
for $h>h_c^{(2)}$. Let $P^{(+)} = | 000 \ket \bra 000| + | \overline{W} \ket \bra
\overline{W}|$ and $P^{(-)} = | 111 \ket \bra 111| + | W \ket \bra W|$ be the
corresponding projection operators. Furthermore, we define $\Sigma_x^{(+)} =
|000\ket \bra \overline{W}| + |\overline{W}\ket \bra 000|$, $\Sigma_y^{(+)} =
-i|000\ket \bra \overline{W}| + i|\overline{W}\ket \bra 000|$, and
$\Sigma_z^{(+)} = |000\ket \bra 000| -|\overline{W}\ket \bra \overline{W} |$,
as well as $\Sigma_k^{(-)} = \sigma_x \otimes \sigma_x \otimes \sigma_x \Sigma_k^{(+)}
\sigma_x \otimes \sigma_x \otimes \sigma_x$, $k=x,y,z$.  This yields
the effective projected two-level ground state Hamiltonian
\begin{eqnarray}
H_{g} (\gamma,h,\phi) = \left\{ \begin{array}{lll}
P^{(+)} H(\gamma,h,\phi) P^{(+)} & = &
\mathcal{E}_0 (\gamma,h) P^{(+)} \\
 & & \\
 & & + \Delta \mathcal{E} (\gamma,h)
\left( \sin \left[ \Theta (\gamma,h) \right] \cos (2\phi) \Sigma_x^{(+)} \right. \\
 & & \\
 & & + \sin \left[ \Theta (\gamma,h) \right] \sin (2\phi) \Sigma_y^{(+)} \\
 & & \\
 & & \left. + \cos \left[ \Theta (\gamma,h) \right] \Sigma_z^{(+)} \right) , \\
 & & \\
 & & h_c^{(1)} < h < h_c^{(2)}, \\
 & & \\
P^{(-)} H(\gamma,h,\phi) P^{(-)} & = &
\mathcal{E}_0 (\gamma,-h) P^{(-)} + \\
 & & \\
 & & \Delta \mathcal{E} (\gamma,-h)
\left( \sin \left[ \Theta (\gamma,-h) \right] \cos (2\phi) \Sigma_x^{(-)} \right. \\
 & & \\
 & & -\sin \left[ \Theta (\gamma,-h) \right] \sin (2\phi) \Sigma_y^{(-)} \\
 & & \\
 & & \left. + \cos \left[ \Theta (\gamma,-h) \right] \Sigma_z^{(-)} \right) , \\
 & & \\
 & & h>h_c^{(2)} . \\
\end{array} \right.
\end{eqnarray}
This describes a spin$-\frac{1}{2}$ particle exposed to an effective
magnetic field with strength $\Delta \mathcal{E} (\gamma,\pm h)$ that
takes the form
\begin{eqnarray}
\boldsymbol{B}^{(\pm)} & = & \Delta \mathcal{E} (\gamma,\pm h)
\left\{ \sin \left[ \Theta (\gamma,\pm h) \right] \cos (2\phi) ,
\pm \sin \left[ \Theta (\gamma,\pm h) \right] \sin (2\phi) ,
\cos \left[ \Theta (\gamma,\pm h) \right] \right\}
\nonumber \\
 & = & \Delta \mathcal{E} (\gamma,\pm h) {\bf n} (\gamma,\pm h,\phi) .
\end{eqnarray}
Here, we have introduced the unit vectors ${\bf n} (\gamma,\pm h,\phi)$ that
rotate with twice the spin rotation angle $\phi$ and make polar angles
$\Theta (\gamma,\pm h)$ with the effective $z$ axis. The sign difference
in the expression for the two types of ground state Berry phases in Eq.
(\ref{eq:purebp}) originates from that $\boldsymbol{B}^{(\pm)}$ rotate
in opposite direction. It is visible that the origin of the nontrivial
ground state Berry phase is a monopole sitting at the point where
$\Delta \mathcal{E} = 0$. This happens at $(\gamma,h)=(1,-\frac{1}{3})$
in the low field regime ($h_c^{(1)} < h < h_c^{(2)}$) and at
$(\gamma,h) =(1,\frac{1}{3})$ in the high field regime ($h>h_c^{(2)}$);
for $\gamma \neq 1$ there is an avoided crossing at
$h=-\frac{1}{6} (1+\gamma)$ ($h=\frac{1}{6} (1+\gamma)$) corresponding
to a minimal energy difference $2\Delta \mathcal{E} = \frac{1}{\sqrt{3}}
|1-\gamma|$ in the low (high) field regime. The Berry effective gauge
field takes the magnetic monopole form
\begin{eqnarray}
\boldsymbol{B}_{{\textrm{eff}}}^{(\pm)} =
\mp \frac{1}{2} \frac{{\bf n} (\gamma,\pm h,\phi)}
{\left[ \Delta \mathcal{E} (\gamma,\pm h) \right]^2}
\end{eqnarray}
and the Berry phase shown in Fig. \ref{fig:bp} is the flux of
$\boldsymbol{B}_{{\textrm{eff}}}^{(\pm)}$ through any surface enclosed
by the curve traversed in parameter space $(\Delta \mathcal{E},\Theta,2\phi)$,
where $\Delta \mathcal{E}$ and $\Theta$ are determined by $\gamma$ and h.
We may therefore interpret the jump across the crossing line
$\gamma \mapsto h_c^{(2)}$ as an interplay between a jump
in polar angle $\Theta$ and that the two types of ground state feel
monopoles sitting at different points in the $(\gamma,h)$ plane.

\subsection{Subsystems}
An interferometer experiment to detect the Berry phase could be
set up for one or two of the qubits. As the states of the subsystems
in general are mixed, the corresponding Berry phases would coincide
with the mixed state geometric phase in Ref. \cite{sjoqvist00b}, applied
to adiabatic evolution. Here, we examine the behavior
of these mixed state Berry phases in the LMG system.

We calculate the subsystem Berry phases under slow rotation around
the $z$ axis. To this end, we need the reduced ground states
$\rho^{(\pm)}$ and $\varrho^{(\pm)}$ of the one- and two-qubit subsystem,
respectively. Taking into account the translational symmetry of the
ground states $V_{-}^{(\pm)}$, these marginal states for any of the
qubit or qubit pair read
\begin{eqnarray}
\rho^{(\pm)} = \frac{1}{2} \left[ \hat{1} + r(\gamma,\pm h) \sigma_z \right]
\end{eqnarray}
and
\begin{eqnarray}
\varrho^{(\pm)} =
\frac{1}{2} \left[ 1+r(\gamma,\pm h) \right] |\psi_1^{(\pm)} \ket \bra \psi_1^{(\pm)}| +
\frac{1}{2} \left[ 1-r(\gamma,\pm h) \right] |\psi_2^{(\pm)} \ket \bra \psi_2^{(\pm)}|
\label{eq:2qubit}
\end{eqnarray}
respectively. Here,
\begin{eqnarray}
r(\gamma,h) & = & \frac{1}{3} \left[ 1+2\cos \Theta (\gamma,h) \right] ,
\nonumber \\
|\psi_{1}^{(+)} \ket & = &
\frac{1}{\sqrt{2 + \cos \left[ \Theta (\gamma,h) \right]}}
\left\{ \sqrt{3} \cos \left[ \frac{1}{2} \Theta (\gamma,h) \right]
|00\ket +
\sin \left[ \frac{1}{2} \Theta (\gamma,h) \right] |11\ket \right\} ,
\nonumber \\
|\psi_{1}^{(-)} \ket & = &
\frac{1}{\sqrt{2 + \cos \left[ \Theta (\gamma,-h) \right]}}
\left\{ \sqrt{3} \cos \left[ \frac{1}{2} \Theta (\gamma,-h) \right]
|11\ket +
\sin \left[ \frac{1}{2} \Theta (\gamma,-h) \right] |00\ket \right\} ,
\nonumber \\
|\psi_{2}^{(\pm)} \ket & = &
\frac{1}{\sqrt{2}} \Big( |01\ket + |10\ket \Big) ,
\label{eq:eigenvaluesvectors}
\end{eqnarray}
which define the eigenvectors $ e^{-i\frac{1}{2} \phi (\sigma_z \otimes
\hat{1} + \hat{1} \otimes \sigma_z)} |\psi_{\mu}^{(\pm)} \ket$ of
$\varrho^{(\pm)} (\phi)$ corresponding to its nonzero eigenvalues.

We notice that the one-qubit Berry phases vanish since the corresponding
reduced density operators are diagonal in the $|0\ket,|1\ket$ basis and
thereby commute with $e^{-i\frac{1}{2} \sigma_z}$. On the other hand, the
two-qubit Berry phases may be non-vanishing. To see this, we note that
the reduced two-qubit density operator reads $\varrho^{(\pm)} (\phi) =
e^{-i\frac{1}{2} \phi (\sigma_z \otimes \hat{1} + \hat{1} \otimes \sigma_z)}
\varrho^{(\pm)} e^{i\frac{1}{2} \phi (\sigma_z \otimes \hat{1} +
\hat{1} \otimes \sigma_z)}$. We see that $\varrho^{(\pm)} (\phi) \neq
\varrho^{(\pm)}$, which opens up for nontrivial mixed state Berry phase
$\Gamma^{(\pm)}$ of the two-qubit states.

\begin{figure}
\includegraphics{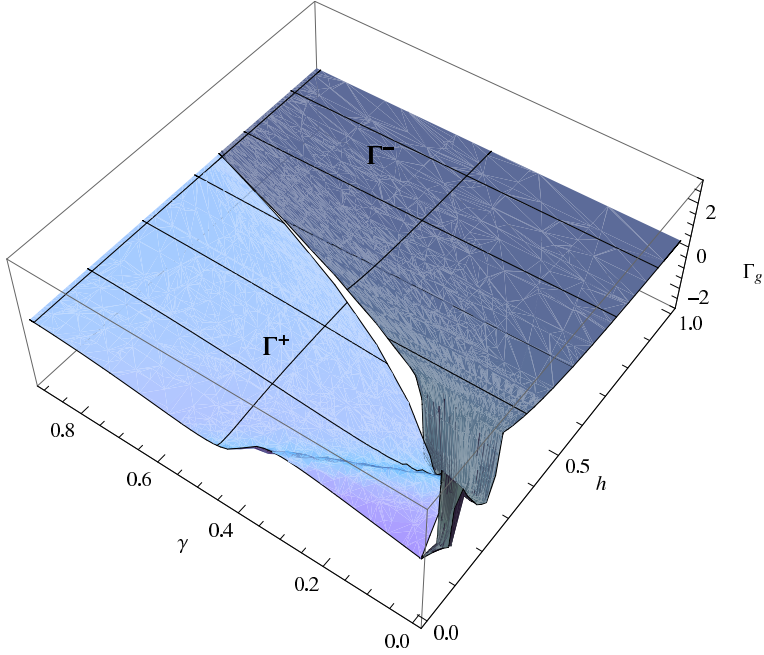}
\caption{\label{fig:bp1} The absolute value $|\Gamma_g|$ of the two-qubit
Berry phase $\Gamma_g$ of the reduced ground state as a function of the isotropy parameter
$\gamma$ and the magnetic field strength $h$.
(This figure is in colour only in the electronic version)}
\end{figure}

By applying \cite{sjoqvist00b} to a cyclic adiabatic evolution, we obtain
\begin{eqnarray}
\Gamma^{(\pm)} =
\arg \left( \sum_{\mu} p_{\mu}^{(\pm)} e^{i\beta_{\mu}^{(\pm)}} \right) ,
\label{eq:generalmsgp}
\end{eqnarray}
where $p_{\mu}^{(\pm)}$ and $\beta_{\mu}^{(\pm)}$ are the density operator's
eigenvalues and eigenstate Berry phases, respectively.
For cyclic evolution, we obtain from Eq. (\ref{eq:generalmsgp}) the two-qubit
geometric phase $\Gamma_g$ of the ground state as
\begin{eqnarray}
\Gamma_g = \left\{
\begin{array}{lll}
\Gamma^{(+)} & = & \arg \left\{ \left( 2 +
\cos \left[ \Theta (\gamma,h) \right] \right) e^{i\beta^{(+)}_{1}} +
\left( 1 - \cos \left[ \Theta (\gamma,h) \right] \right)
e^{i\beta^{(+)}_{2}} \right\} \\
 & & \\
 & = & \arctan \left( \frac{ \left( 2 +
\cos \left[ \Theta (\gamma,h) \right] \right)
\sin \beta^{(+)}_{1}}{1 - \cos \left[ \Theta (\gamma,h) \right] +
\left( 2 + \cos \left[ \Theta (\gamma,h) \right] \right)
\cos \beta^{(+)}_{1}} \right) , \ \ h_c^{(1)} < h < h_c^{(2)} \\
 & & \\
\Gamma^{(-)} & = & \arg \left\{ \left( 2 +
\cos \left[ \Theta (\gamma,-h) \right] \right) e^{i\beta^{(-)}_{1}} +
\left( 1 - \cos \left[ \Theta (\gamma,-h) \right] \right)
e^{i\beta^{(-)}_{2}} \right\} \\
 & & \\
 & = & \arctan \left( \frac{ \left( 2 +
\cos \left[ \Theta (\gamma,-h) \right] \right)
\sin \beta^{(-)}_{1}}{1 - \cos \left[ \Theta (\gamma,-h) \right] +
\left( 2 + \cos \left[ \Theta (\gamma,-h) \right] \right)
\cos \beta^{(-)}_{1}} \right) , \ \ h>h_c^{(2)}
\end{array} \right.
\end{eqnarray}
where we have used that $\beta^{(\pm)}_{2} = 0$ since $|\psi_1^{(\pm)}\ket$
are eigenvectors of $\sigma_z \otimes \hat{1} + \hat{1} \otimes \sigma_z$.
Here,
\begin{eqnarray}
\beta^{(\pm)}_{1} =
\mp 2\pi \frac{1-\cos \left[ \Theta (\gamma,\pm h) \right]}
{2 + \cos \left[ \Theta (\gamma,\pm h) \right]} ,
\end{eqnarray}
which is $\beta^{(\pm)}$ quenched by a factor $\left\{ 2 +
\cos \left[ \Theta (\gamma,\pm h) \right] \right\}^{-1}$.
The absolute value $|\Gamma_g|$ of the two-qubit Berry phase $\Gamma_g$ of
the ground state is shown in Fig. \ref{fig:bp1}.

Note that the relative phase $\arg \Tr \left( e^{-i\frac{1}{2} \phi
(\sigma_z \otimes \hat{1} + \hat{1} \otimes \sigma_z)} \varrho^{(\pm)} \right)$
\cite{sjoqvist00b} between the initial and final two-qubit states is measurable
for all $(\gamma,h)$, as the visibility $\left| \Tr \left( e^{-i\frac{1}{2}
\phi (\sigma_z \otimes \hat{1} + \hat{1} \otimes \sigma_z)} \varrho^{(\pm)}
\right) \right|$ cannot vanish for these states. On the other hand, the
geometric part $\Gamma_g$ of this relative phase is not always well-defined;
in fact, it is undefined if $\varrho^{(\pm)}$ has nonzero
degenerate eigenvalues \cite{sjoqvist00b}, i.e., when $\cos \Theta =
- \frac{1}{2}$. Thus, the mixed state Berry phase is undefined precisely
when the three-qubit ground state is of GHZ form, i.e., at $(\gamma,h)=(0,0)$, see
Eq. (\ref{eq:gsorigin}). Note also the oscillatory behavior of $\Gamma_g$
that is visibility in Fig. \ref{fig:bp1} when approaching the degeneracy.
These oscillations may be understood from the fact that the mixed state
Berry phase in unitary evolution is known to vary more rapidly in the
vicinity of a degeneracy point of the corresponding density operator
\cite{bhandari02}. Finally, just as the three-qubit Berry
phase factor, the two-qubit Berry phase factor is smooth across the crossing
point at $h=\frac{2}{3}$ in the isotropic LMG model. Indeed, $\Gamma_g$
vanishes for all $h$ as $\beta_1^{(\pm)}$ is an integer multiple of $2\pi$
when $\gamma = 1$.

\section{Fidelity}
\label{sec:fidelity}

\subsection{Fidelity susceptibility}
Fidelity is a measure of similarity between different quantum states and
is therefore expected to be sensitive to abrupt changes in the ground state
properties in many-body systems. This has triggered work to use fidelity
measures in the context of quantum critical phenomena
\cite{zanardi06,zanardi07,venuti07,you07,chen08,tribedi08}. Here, we
analyze the fidelity susceptibility \cite{you07} in the present
three-qubit LMG system.

\begin{figure}[htb]
\includegraphics[width=8cm]{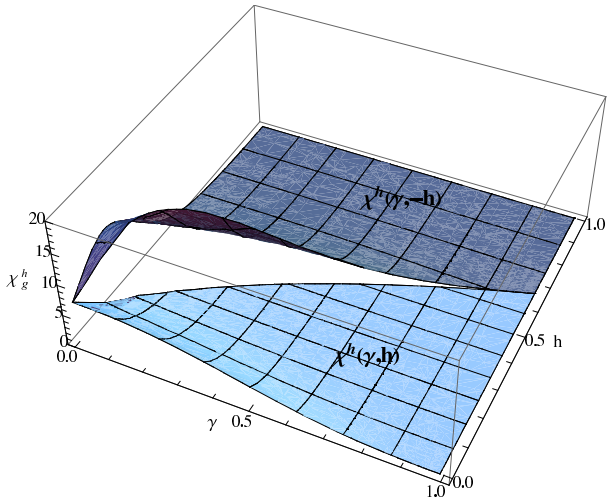}
\caption{\label{fig:FS} Fidelity susceptibility $\chi_g^h$ of ground
state as a function of the isotropy parameter $\gamma$ and the magnetic field
strength $h$. (This figure is in colour only in the electronic version)}
\end{figure}

An abrupt change in the LMG model system can be induced by slowly tuning the external
magnetic field $h$ across the crossing value $h_c^{(2)} = \frac{2}{3} \sqrt{\gamma}$
at fixed $\gamma$. The fidelity susceptibility $\chi_g^h (\gamma,h)$ is taken to
measure the response of the ground state to such variations in $h$. A convenient
form for this $\chi_g^h (\gamma,h)$ may be found by writing the LMG Hamiltonian as
\begin{eqnarray}
H (\gamma,h) = H_0 + h H_I ,
\end{eqnarray}
where $H_0 = -\frac{1}{6} \left( S_x^2 + \gamma S_y^2 \right)$ is independent
of $h$ and $H_I = -S_z$ is the driving Hamiltonian, the relative strength of
$H_0$ and $H_I$ being controlled by $h$. Let $|V_g (\gamma,h) \ket = |V_{-}^{(\pm)} \ket$
be the normalized ground state of $H (\gamma,h)$ and $E_g (\gamma,h)$ the
corresponding ground state energy. The fidelity susceptibility $\chi_g^h (\gamma,h)$
of $V_g$ is defined as the leading nontrivial contribution in $\delta h$ to the
fidelity $F_g$ between the ground states $|V_g (\gamma,h) \ket$ and
$|V_g (\gamma,h + \delta h) \ket$, according to
\begin{eqnarray}
F_g (\gamma,h,\delta h) & = &
\left| \langle V_g (\gamma,h) |
V_g (\gamma, h + \delta h) \rangle \right|
\nonumber \\
 & = & 1-\frac{1}{2} \chi_g^h (\gamma,h) \delta h^2
 + \ldots ,
\end{eqnarray}
where we may note that $\chi_g^h$ is independent of the arbitrary parameter
$\delta h$. By expanding to second order in $\delta h$ and using the form of
$H (\gamma,h)$, we obtain \cite{you07}
\begin{eqnarray}
\chi_g^h (\gamma,h) =
\sum_{n\neq g} \frac{\left| \bra V_n (\gamma,h)| H_I |
V_g (\gamma,h) \ket \right|^2}
{\left[ E_n (\gamma,h) - E_g (\gamma,h) \right]^2} ,
\label{eq:fidsuscept}
\end{eqnarray}
where $|V_n (\gamma,h) \ket$ and $E_n (\gamma,h)$ are eigenvectors
and eigenvalues, respectively, of $H (\gamma,h)$. By inserting Eq.
(\ref{eq:eigensolutions}) into Eq. (\ref{eq:fidsuscept}) and using
$H_I = -S_z$, we obtain
\begin{eqnarray}
\chi_g^h (\gamma,h) = \left\{ \begin{array}{lll}
\chi (\gamma,h) & = & \frac{\left| \bra V_{+}^{(+)}| (-S_z) |V_{-}^{(+)} \ket
\right|^2}{\left[ \Delta \mathcal{E} (\gamma,h) \right]^2} =
\frac{\sin^2 \left[ \Theta (\gamma,h) \right]}
{\left[ \Delta \mathcal{E} (\gamma,h) \right]^2}, \ \ h_c^{(1)} < h < h_c^{(2)} , \\
 & & \\
\chi  (\gamma,-h) & = & \frac{\left| \bra V_{+}^{(-)}| (-S_z) |V_{-}^{(-)} \ket
\right|^2}{\left[ \Delta \mathcal{E} (\gamma,-h) \right]^2} =
\frac{\sin^2 \left[ \Theta (\gamma,-h) \right]}
{\left[ \Delta \mathcal{E} (\gamma,-h) \right]^2}, \ \ h > h_c^{(2)} .
\end{array} \right.
\end{eqnarray}
We may note that $S_z$ does not couple the degenerate states $V_1^{(\pm)}$ and
$V_2^{(\pm)}$ to any of the two candidate ground states, since $V_{1,2}^{(\pm)}$
and $V_{-}^{(\pm)}$ belong to different $S^2$ eigenvalues. The fidelity susceptibility
$\chi_g^h (\gamma,h)$ is shown in Fig. \ref{fig:FS}.

The fidelity susceptibility of the LMG ground state vanishes identically for
$\gamma = 1$ as $\Theta (1,h) = 0$ or $\pi$. $\chi_g^h$ therefore shares the
behavior of the Berry phase factors $e^{i\beta_g}$
and $e^{i\Gamma_g}$ in that it is smooth across the crossing point in the
isotropic ($\gamma = 1$) case. Furthermore, $\chi_g^h (\gamma,h_c^{(2)-})$
decreases monotonically towards zero as a function of $\gamma$ when $\gamma$
increases. This can be explained by noting that $\sin^2 ( \Theta )$ decreases
monotonically when $\Theta$ increases from $\Theta (0,0) = \frac{2\pi}{3}$ to
$\Theta (1,h_c^{(2)-}) = \pi$. On the other hand, $\chi_g^h (\gamma,h_c^{(2)+})$
has a local maximum since to decrease $\Theta (0,0) = \frac{2\pi}{3}$ to
$\Theta (1,h_c^{(2)+}) = \pi$ one must pass the intermediate angle $\Theta =
\frac{\pi}{2}$ at which $\sin \Theta$ has its maximum. Furthermore, we note
that the fidelity susceptibility is singular close to the degeneracies
at $(\gamma,h)=(1,\pm \frac{1}{3})$ which corresponds to the locations
of the effective magnetic monopoles and where the adiabatic approximation
breaks down. The singular behavior expresses the fact that small variations
in the parameters may cause transitions between the two orthogonal states that
cross at these points.

\subsection{Partial-state fidelity susceptibility}
Partial-state fidelity susceptibility has been developed to deal with
the response of a subsystem $s$ to the driving Hamiltonian \cite{paunkovic08a,kwok08}.
It is defined as the leading nontrivial contribution of the Bures-Uhlmann
fidelity $F_{s;g}$ \cite{bures69,uhlmann76} of two marginal
ground states $\rho_g (\gamma,h) = \Tr_p | V_g (\gamma,h) \ket
\bra V_g (\gamma,h)|$ and $\rho_g (\gamma,h + \delta h) =
\Tr_p |\psi_g (\gamma,h + \delta h) \ket \bra \psi_g (\gamma,h +
\delta h)|$, $\Tr_p$ being partial trace over one or two of the qubits.
Explicitly,
\begin{eqnarray}
F_{s;g} (\gamma,h,\delta h) & = &
\Tr \sqrt{\sqrt{\rho_g (\gamma,h)} \rho_g (\gamma,h+\delta h)
\sqrt{\rho_g (\gamma,h)}}
\nonumber \\
 & = & 1-\frac{1}{2} \chi_{s;g}^h (\gamma,h)
 \delta h^2 + \ldots
\end{eqnarray}
defines the partial-state susceptibility $\chi_{s;g}^h$ of the energetic
ground state \cite{paunkovic08a,kwok08}.

\begin{figure}[htb]
\includegraphics[width=8cm]{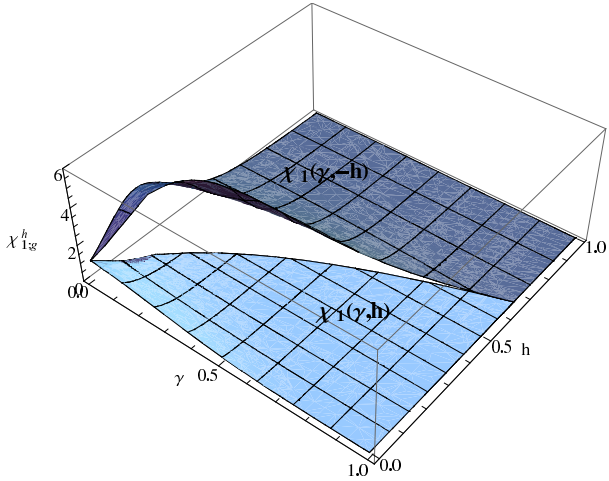}
\caption{\label{fig:ps1} One-qubit partial-state fidelity susceptibility
$\chi_{1;g}^h$
of the LMG ground state as a function of the isotropy parameter $\gamma$
and the magnetic field strength $h$.
(This figure is in colour only in the electronic version)}
\end{figure}

Let us first consider the one-qubit partial-state fidelity $\chi_{1;g}^h$
with respect to variations of $h$. The marginal ground state $\rho^{(\pm)}$
of any of the three qubits is diagonal in the fixed $|0\ket,|1\ket$ basis.
This implies that only changes in the purity parameter $r(\gamma,h)$
contribute to $\chi_{1;g}^h$. An explicit calculation yields
\begin{eqnarray}
\chi_{1;g}^h (\gamma,h)  = \left\{ \begin{array}{lll}
\chi_{1} (\gamma,h) & = &
\frac{1}{2} \frac{1}{1-[r (\gamma,h)]^2}
\left[ \frac{\partial}{\partial h} r (\gamma,h) \right]^2 \\
 & & \\
 & = & \frac{\sin^2 \left[ \Theta (\gamma,h) \right]}
{4 - 2\cos \left[ \Theta (\gamma,h) \right] -
2\cos^2 \left[ \Theta (\gamma,h) \right]}
\left[ \frac{\partial}{\partial h} \Theta (\gamma,h) \right]^2 , \\
 & & \\
 & & h_c^{(1)} < h < h_c^{(2)} , \\
 & & \\
\chi_{1} (\gamma,-h) & = &
\frac{1}{2} \frac{1}{1-[r (\gamma,-h)]^2}
\left[ \frac{\partial}{\partial h} r (\gamma,-h) \right]^2 \\
 & & \\
 & = & \frac{\sin^2 \left[ \Theta (\gamma,-h) \right]}
{4 - 2\cos \left[ \Theta (\gamma,-h) \right] -
2\cos^2 \left[ \Theta (\gamma,-h) \right]}
\left[ \frac{\partial}{\partial h} \Theta (\gamma,-h) \right]^2 , \\
 & & \\
 & & h > h_c^{(2)} .
\end{array} \right.
\end{eqnarray}
The one-qubit partial-state fidelity susceptibility $\chi_g^h (\gamma,h)$
is shown in Fig. \ref{fig:ps1}.

\begin{figure}[htb]
\includegraphics[width=8cm]{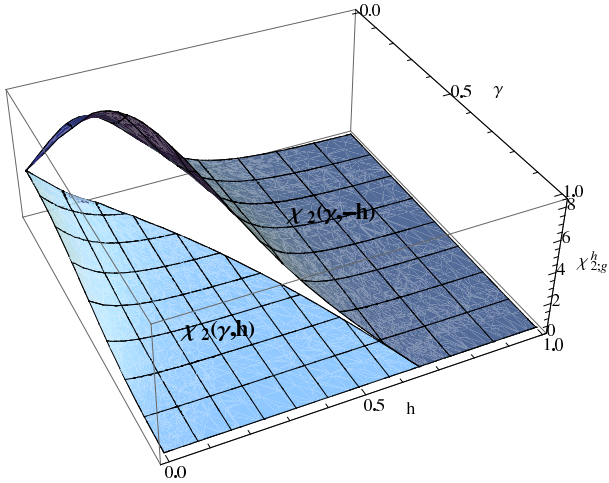}
\caption{\label{fig:ps2} Two-qubit partial-state fidelity
susceptibility $\chi_{2;g}^h$ of the LMG ground state as a
function of the isotropy parameter $\gamma$ and the magnetic field
strength $h$. (This figure is in colour only in the electronic version)}
\end{figure}

The two-qubit partial-state fidelity $\chi_{2;g}^h$ originates from
changes in the purity of $\varrho^{(\pm)}$ and in the parameter
dependent eigenvector $|\psi_1^{(\pm)} \ket$ of $\varrho^{(\pm)}$. The
relevant purity parameter $r(\gamma,h)$, which implies that the
contribution to $\chi_{2;g}^h$ from the purity coincides
with the one-qubit partial-state fidelity susceptibility $\chi_{1;g}^h$.
The additional contribution related to the change of $|\psi_1^{(\pm)} \ket$ equals
the corresponding pure state fidelity susceptibility, weighted by the probability
$\frac{1}{2} \left[ 1 + r(\gamma,\pm h) \right]$. Explicitly, we have
\begin{eqnarray}
\chi_{2;g}^h (\gamma,h)  = \left\{ \begin{array}{lll}
\chi_2 (\gamma,h) & = &
\frac{1}{2} \frac{1}{1-[r (\gamma,h)]^2}
\left[ \frac{\partial}{\partial h} r (\gamma,h) \right]^2 \\
 & & \\
 & & + \left[ 1+r(\gamma,h) \right] \left( \bra \frac{\partial}{\partial h}
\psi_1^{(+)} |\frac{\partial}{\partial h} \psi_1^{(+)} \ket \right. \\
 & & \\
 & & \left. - \bra \frac{\partial}{\partial h} \psi_1^{(+)} |\psi_1^{(+)} \ket
\bra \psi_1^{(+)} |\frac{\partial}{\partial h} \psi_1^{(+)} \ket \right) \\
 & & \\
 & = & \chi_1 (\gamma,h) +
\frac{1}{2+\cos \left[\Theta (\gamma,h)\right]}
\left[ \frac{\partial}{\partial h} \Theta (\gamma,h) \right]^2 , \\
 & & \\
 & & h_c^{(1)} < h < h_c^{(2)} , \\
 & & \\
\chi_2 (\gamma,-h) & = &
\frac{1}{2} \frac{1}{1-[r (\gamma,-h)]^2}
\left[ \frac{\partial}{\partial h} r (\gamma,-h) \right]^2 \\
 & & \\
 & & + \left[ 1+r (\gamma,-h) \right] \left( \bra \frac{\partial}{\partial h}
\psi_1^{(-)} |\frac{\partial}{\partial h} \psi_1^{(-)} \ket \right. \\
 & & \\
 & & \left. - \bra \frac{\partial}{\partial h} \psi_1^{(-)} |\psi_1^{(-)} \ket
\bra \psi_1^{(-)}|\frac{\partial}{\partial h} \psi_1^{(-)} \ket \right) \\
 & & \\
 & = & \chi_1 (\gamma,-h) +
\frac{1}{2+\cos \left[\Theta (\gamma,-h)\right]}
\left[ \frac{\partial}{\partial h} \Theta (\gamma,-h) \right]^2 , \\
 & & \\
 & & h > h_c^{(2)} .
\end{array} \right.
\label{eq:2fidsuscept}
\end{eqnarray}
The two-qubit partial-state fidelity susceptibility $\chi_{2;g}^h (\gamma,h)$
is shown in Fig. \ref{fig:ps2}.

Both the one- and two-qubit partial-state fidelity susceptibilities behave
similarly as that of the full ground state $V_g$: both $\chi_{1;g}^h
(\gamma,h)$ and $\chi_{2;g}^h (\gamma,h)$ vanish in the isotropic
($\gamma = 1$) case and there is a similar dependence on $\gamma$ close to the
crossing line $\gamma \mapsto h_c^{(2)}$. Furthermore, by comparing Figs.
\ref{fig:FS}, \ref{fig:ps1}, and \ref{fig:ps2}, we notice that the pure state
fidelity susceptibilities is typically larger than the partial-state fidelity
susceptibilities. It is apparent that $\chi_{2;g}^h (\gamma,h) \geq
\chi_{1;g}^h (\gamma,h)$ since the second term in the right-hand side
of Eq. (\ref{eq:2fidsuscept}). This may be interpreted to be a consequence of the
loss of purity for each qubit that is traced out.

\section{Conclusions}
A detailed characterization of the ground state of a three-qubit
Lipkin-Meshkov-Glick (LMG) type model has been given. We have
calculated Berry phases for the three-qubit state as well as
for the reduced two-qubit state in the case of local, coordinated
$2\pi$ rotation around the axis of the external magnetic field.
We have identified an underlying
two-level structure of the three-qubit Berry phase and found the
relevant magnetic monopole distribution. The energetic ground state of
the model is of GHZ-type if the external field and the isotropy
parameter both vanish. The reduced two-qubit state at this point
in parameter space is two-fold degenerate and separable, from which
follows that the corresponding mixed state Berry phase is undefined.
The three- and two-qubit Berry phases vanish modulus $2\pi$ in the
isotropic LMG model.

We have calculated the fidelity susceptibility and the one-
and two-qubit partial-state fidelity susceptibility for the LMG model.
These fidelity susceptibilities all behave similarly, but decreases
in size for each qubit being traced out. We have found
that the fidelity susceptibilities all vanish in the isotropic LMG model.
Analogously to the fidelity susceptibility and Berry phase in the pure
state case, the partial-state fidelity susceptibility and the Uhlmann
holonomy \cite{uhlmann86} measure the geometry of the space
of mixed quantum states. This observation makes it natural to ask
whether the Uhlmann holonomy may yield further insights into the
ground state properties of interacting spin models. Paunkovi\'{c} and
Rocha Vieira \cite{paunkovic08b} have found a rich structure in the
Uhlmann holonomy for thermal states in the Stoner-Hubbard and BCS models.
A similar calculation of the partial-state holonomy seems pertinent in
relation to the present work.

We hope that the analysis presented in this work may trigger investigations
of few-qubit models to explore further their effective magnetic monopole
structure and its associated state space geometry.

\section*{Acknowledgments}
E.S. acknowledges financial support from the Swedish Research
Council (VR) R.R. acknowledges financial support from the Norwegian
Research Council. U.B. would like to acknowledge thankfully the
financial support of the Council of Scientific and Industrial
Research, India Grant No. SPM-07/489(0034)/2007.

\end{document}